\documentclass[preprint]{elsarticle}

\usepackage{amssymb}
\usepackage{booktabs} 
\usepackage{graphicx}
\usepackage{xcolor}
\usepackage{tablefootnote}

\journal{Astroparticle Physics}

\begin{document}

\begin{frontmatter}



\title{Cosmic-ray propagation under consideration of a spatially resolved source distribution}


\author[uibk]{J.~Thaler}
\ead{j.thaler@uibk.ac.at}

\author[uibk]{R.~Kissmann}
\author[uibk]{O.~Reimer}
\address[uibk]{Institut f\"ur Astro- und Teilchenphysik, Leopold-Franzens-Universit\"at Innsbruck, A-6020 Innsbruck, Austria}
\begin{abstract}
Cosmic rays (CRs) are an integral component of the interstellar medium, producing broadband emission while interacting with other Galactic matter components like the interstellar gas or magnetic fields. In addition to observations, numerical simulations of CR propagation through the Galaxy help to increase the level of understanding of Galactic CR transport and diffuse $\gamma$-ray emission as seen by different experiments. Up to now, the standard approach at modelling source distributions used as input for such transport simulations often rely on radial symmetry and analytical functions rather than individual, observation-based sources. We aim at a redefinition of existing CR source distributions by combining sources observed with the H.E.S.S. experiment and simulated random sources, which follow the matter density in the Milky Way.  As a result, H.E.S.S.-inspired Galactic CR source distributions are inferred.
\newline
We use the PICARD code to perform 3D-simulations of nuclei and electrons in CR propagation using our hybrid source distribution models. Furthermore, also gamma-ray maps and spectra, simulated with the redefined source models, are evaluated in different regions in the Galaxy and compared with each other to determine the statistical scatter of the underlying distributions. We find global consistency between our models and in comparison to previous simulations, with only some localised fluctuations, e.g. in the spiral arms.
\newline
This implementation of a three-dimensional source model based on observations and simulations enables a new quality of propagation modelling. It offers possibilities for more realistic CR transport scenarios beyond radial symmetry and delivers meaningful results in both the arm and interarm regions of the Galaxy. This gives a more realistic picture of the Galactic $\gamma$-ray sky by including structures from the source model and not just the gas distributions. 

\end{abstract}



\end{frontmatter}


\section{Introduction}
The non-thermal emission from CR particles interacting with other components of the interstellar medium, like gas, radiation and magnetic fields dominates Galactic emission over a vast range of wavelengths.

However, the origin of Galactic CRs poses a riddle to science since their discovery in 1912 \citep{hess_uber_1912}. Nowadays astrophysical objects such as supernova remnants \citep{ackermann_detection_2013} or pulsar wind nebulae \citep{aharonian_high_2008} are considered the main Galactic candidates for particle acceleration but also other objects like, for instance, $\gamma$-ray binaries \citep{DUBUS2015661} are possibly contributing sources. Whilst propagating through the Galaxy from their putative sources to the observer, CRs interact with radiation fields and particles from the interstellar matter, producing diffuse $\gamma$-rays.
Due to the CR background the measurement of this diffuse $\gamma$-ray flux at very high energies is difficult for ground-based telescopes, but there have been attempts to file the source distribution.
The most thorough survey of CR sources realized at high GeV to TeV energies was the High Energy Stereoscopic System's (H.E.S.S.) Galactic Plane Survey  \citep{hess_collaboration_hess_2018}, which was carried out from 2004 to 2013. Hereby, the Galactic plane was observed using a data set of about $2700$ hours of observations. This led to a new, very high-energy $\gamma$-ray source catalogue and resolved $\gamma$-ray skymaps of the inner Milky Way, more precisely the region containing the majority of Galactic $\gamma$-ray sources \citep{Ackermann_2016}. However, still, only the brightest and closest $\gamma$-ray sources are detected, which presumably leaves a large fraction of the overall source population hidden under the detection threshold of H.E.S.S. or entangled with very faint diffuse emission, stemming from the interactions of propagating CRs.  Therefore,  since not all CR sources can be detected with current instruments, population synthesis is needed. This technique dates back to the early 1970s \citep{2003MNRAS.344.1000B}, is actively used for example by \citep{Strong:2006hf} to characterize the whole Galactic CR source population and was usually used to build continuous, mostly axisymmetric analytical source distributions. In \citep{egberts_unresolved_2017} generic 3D source populations are created in order to model a typical source distribution for the whole Galaxy with the help of synthetic sources. On one hand this approach is able to predict the source count to be detected by future instruments but on the other hand there are many assumptions inherent in such simulations, for example morphological aspects such as the geometry of the Milky Way. Therefore, a promising way to get a more complete and authentic image of our Galaxy is the combination of observed and synthetic sources. It is the goal of this work to use this approach to improve existing models of the Galactic CR source distribution.  Thereby, we want to complement the observed source sample \citep{hess_collaboration_hess_2018} with simulated sources \citep{steppa_constantin_model_nodate} to create a source population model of the whole Galaxy.  This approach is necessary because at TeV energies, where we put our focus on, momentarily unresolved sources are assumed to contribute to the diffuse emission in the Galaxy as well as to the direct $\gamma$-ray emission, which still lies below the detection threshold of current instruments. Such a hybrid source population can further be used as input for CR transport simulations using propagation codes like GALPROP \citep{strong_propagation_1998}, DRAGON \citep{maccione_evoli_gaggero_dragon_2011}, and PICARD \citep{kissmann_picard_2014}, where the latter will be used in this work. 
\newline
In this manuscript we will briefly illustrate the construction of our model and discuss its Galactocentric radial distribution. The simulation set-up is introduced and the details about the used propagation parameters are explained. We present results of our simulations on CR  density distributions and spectra as well as $\gamma$-ray emission. These results and corresponding new findings are then discussed critically and the advantages as well as limitations of the new hybrid model are shown. A conclusion and an outlook for future applicability and improvements conclude this paper.
\section{Construction of the new hybrid model}\label{sec:constr_mod}
Many current Galactic source models used in CR propagation codes are limited to axial-symmetry (see e.g. \citep{Green_2015} or \citep{yusifov_revisiting_2004}) and rely solely on analytical models. With the capabilities of the CR-propagation code PICARD \citep{kissmann_picard_2014}, we aim to achieve a new state-of-the-art CR source distribution based on observed as well as simulated sources. Thereby, we assume that $\gamma$-ray sources are also CR sources and in particular that the CR intensity corresponds to the $\gamma$-ray emissivity of the source. As observed sources, we use those found in the H.E.S.S. Galactic Plane survey (HGPS) \citep{hess_collaboration_hess_2018} as well as four more, previously identified, additional H.E.S.S. sources, namely the Galactic center, SN 1006, HESS J0632+057 and the Crab Nebula. To date, the HGPS offers the most comprehensive view on very-high-energy $\gamma$-ray sources in our Galaxy. However, the resulting source catalogue presumably only represents a small fraction of the total source population, because of the limited sensitivity of the HGPS and its incomplete sky coverage. Furthermore, only $31$ of the provided $78$ sources have been firmly identified using additional evidence such as multi-wavelength variability and morphology to reinforce the spatial associations.  Only those firmly identified sources feature distance estimates and can be included in the construction of our model directly.  For the remaining $47$ HGPS sources we need to find a counterpart from the simulated source sample \citep{steppa_constantin_model_nodate} which assigns a distance to each source. To identify a matching simulated source for each of the HGPS sources without firm identification we first find a list of all simulated sources that match each HGPS source in terms of different criteria. One criterion is the source extension where we consider all simulated sources with an extension from $70\%$ to $140\%$ of the size of the observed source in question.  Another criterion we apply to the sources with adequate extensions is the compatibility of the sources in terms of position. Therefore, we calculate the angular separation between the observed HGPS source and all simulated sources with suitable sizes. The upper limit for the simulated sources to be treated as a possible counterpart for the observed source is an angular separation of $0.1^{\circ}$. Thus, we get a list of possible simulated candidates to act as a counterpart for each HGPS source without distance estimates. From this list we choose the best fitting simulated source for each observed source according to its flux by calculating the ratio between the observed flux of the HGPS source and the fluxes of all suitable simulated sources. The simulated source where this ratio is closest to one is selected as the counterpart to the HGPS source and its luminosity is rescaled to fully correspond to its observed standard. 

This leads to a source sample of $82$ observed sources to start the simulated source model with. However, to get a realistic Galactic source model, we still need to include synthetic sources into our simulated source model. The synthetic source model we use is taken from \citep{steppa_constantin_model_nodate}, where multiple simulated source samples are generated. The geometry used in the simulated sample we use here is based on a four-arm spiral Galaxy model \citep{steiman-cameron_cobe_2010}. This is motivated by the observation that Supernova remnants (SNR) and pulsar wind nebulae (PWN), which are thought to be the dominant Galactic CR source candidates, are remnants of massive stars and those stars were born in star formation regions, which predominantly occur in the spiral arms of the Milky Way. The luminosity randomly assigned to each source in the simulation uses a luminosity distribution based on observed source properties of the data set of the HGPS and corrected for the observational bias induced by the sensitivity, as described in more detail in \citep{steppa_constantin_model_nodate}. Thereby, to get a large statistical sample $2493000$ simulated sources have been randomly generated in the Galaxy. 

Since the detected HGPS sources are directly included in our source model, we want to simulate only such additional sources that would not have been observed by H.E.S.S., as we assume to have detected all sources with sufficiently high luminosities and suitable source extensions in the HGPS observation region. Therefore, the simulated source sample is divided into two sub-samples, namely the sources inside and outside the field of view of the HGPS.  For the HGPS the central part of the Milky Way, covering the region from Galactic longitudes of $l=75^{\circ}$ to $0^{\circ}$  as well as from $360^{\circ}$ to $250^{\circ}$ and latitudes of $b=\pm 5^{\circ}$, was scanned.  Accordingly, all simulated sources inside this region are considered inside the field of view of the HGPS. Those sources are further analysed in terms of their flux in comparison to the HGPS sensitivity in the direction of the source. For this, also the radial source extension $\alpha$ of each source needs to be taken into account.The sensitivity $s$ of the HGPS is generally given for point-like sources and, since most sources are extended, we need to scale the sensitivity for each direction according to the radial extension of the source found there \citep{hess_collaboration_hess_2018}.  This scaling is done by convolving the size of the effective point spread function corresponding to the HGPS survey counts maps $\sigma$ with the radial extension of the source $\alpha$ leading to
\begin{equation}
s_{\mathrm{scaled}}=s\cdot \sqrt{\sigma^2+\alpha^2}
\end{equation}
for the corresponding sensitivity. Sources with radial extensions $\alpha > 2^{\circ}$ are considered for our simulated source model without further analysis because they can not be detected by the HGPS due to the background estimation inside of the field of view \citep{hess_collaboration_hess_2018}. For the less extended sources, the obtained local sensitivities $s_{\mathrm{scaled}}$ for each source are compared to the gamma-ray fluxes $F$ the simulated sources would produce at Earth.  All sources for which $F<s_{\mathrm{scaled}}$ could not have been detected in the HGPS and, therefore, are considered for the simulated source model. Furthermore, we also consider the sources outside the field of view of the HGPS for our simulated source model.

Since the number of these possible model sources outnumbers the observed sources in the HGPS by several orders of magnitude, we reduce the number of simulated sources by defining an upper limit for their total luminosity as
\begin{equation}
L_{\mathrm{sim,max}}= L_{\mathrm{obs}}\cdot \frac{ L_{\mathrm{invis}}}{ L_{\mathrm{vis}}}\cdot \frac{N_{\mathrm{FOV/outside FOV}}}{N_{\mathrm{total}}}.
\label{eq:Lum_max}
\end{equation}
Thereby, $L_{\mathrm{obs}}$ is sum of the luminosities of all 78 sources observed in the HGPS and the four previously detected H.E.S.S. sources.
$L_{\mathrm{invis}}$ is the sum of the luminosities of the simulated sources outside the field of view of the HGPS, those inside the field of view with fluxes smaller than the sensitivity in their direction and those with extensions bigger than $2^{\circ}$. Moreover,$L_{\mathrm{vis}}$ is the sum of the luminosities of the simulated sources inside the field of view detectable in the HGPS. The last term of equation \ref{eq:Lum_max} changes depending on whether we want to define the upper limit for the luminosities inside or outside the field of view of the HGPS. In the first case the numerator is $N_{\mathrm{FOV}}$, which represents the number of simulated sources inside the field of view and the denominator $N_{\mathrm{total}}$ is the total number of sources in the simulated sample. In the second case the numerator changes to $N_{\mathrm{outsideFOV}}$, representing the simulated sources outside the HGPS field of view and the denominator stays the same $N_{\mathrm{total}}$. 

This limit for the total luminosity scales the overall observed luminosity with the ratio of the total luminosities inside and outside the field of view. 

Finally, we choose sources randomly from the sample containing sources inside the field of view which are not detectable in the HGPS until we reach the luminosity limit calculated in the first case and, likewise, choose sources outside the field of view until reaching the corresponding luminosity limit calculated in the second case. 

\begin{figure*}
	\centering
		\includegraphics[width=\textwidth]{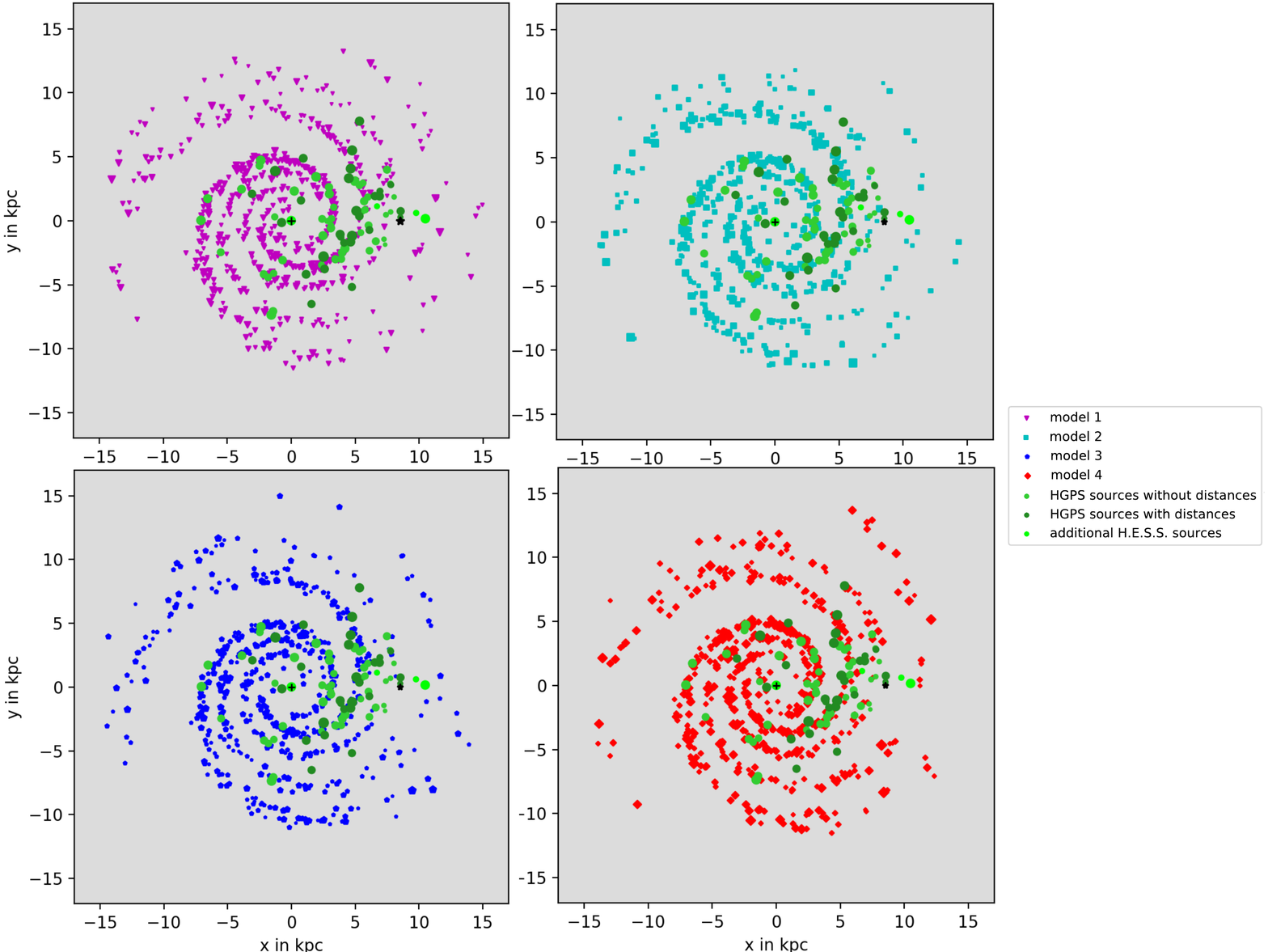}
	\caption{Different random realisations of CR source models (see Table \protect\ref{tab:model_seed_number}) together with the sources from the HGPS: the sources without distance estimate are shown in green, the firmly identified ones with known distance in dark green, and the additional, previously detected sources in light green. Top left: \textit{model 1}, top right: \textit{model 2},  bottom right: \textit{model 3}, bottom left: \textit{model 4}. The size of the dots is scaled by the luminosity of each source, the cross marks the center of the Galaxy and the asterisk the Solar System location, assumed to be at $(8.5,0)~$kpc.}
	\label{fig:source_dist}
\end{figure*}

By producing different random realisations, we find that the number of sources differs in each case. Here, we use four different realisations as shown in table  \ref{tab:model_seed_number} and named \textit{model 1} to \textit{4}, respectively.
\begin{table}
\caption{Different random realisations with the respective total source numbers and label detailed herein.}
\label{tab:model_seed_number}
\begin{center}
\begin{tabular}{lcccc} \toprule
 & \textit{model 1} & \textit{model 2}&\textit{model 3} &\textit{model 4}\\  \midrule
number of sources & $730$ & $671$&$698$&$700$ \\ \bottomrule
\end{tabular}
\end{center}
\end{table}
The total number of sources remains in roughly the same order, namely $700\pm 21$ sources, but the models comprise different morphological distributions.
However, in all distributions tested, the assumed underlying CR source distribution of the Milky Way is clearly visible, as shown in figure \ref{fig:source_dist}. 

\subsection{Radial distribution of the CR sources in the hybrid models}
\begin{figure}
	\centering
		\includegraphics[scale=.5]{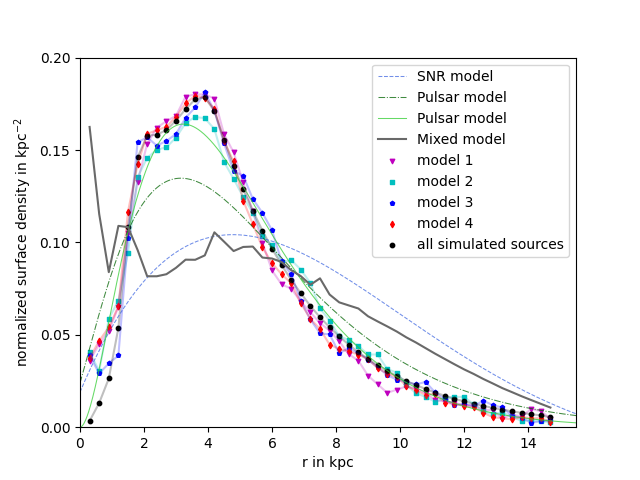}
	\caption{CR source densities as a function of Galactocentric radius. We show distributions for our \textit{models 1-4} and for models from the literature as discussed in the text (top to bottom in the legend: \citep{case_new_1998},\citep{yusifov_revisiting_2004},\citep{lorimer_parkes_2006},\citep{Linden2016}). The black dots represent a distribution containing all sources from the underlying simulated source sample \citep{steppa_constantin_model_nodate}.}
	\label{fig:radial_dist}
\end{figure}

The Galactic distribution of putative CR sources has been studied frequently and by comparison the corresponding Galactocentric radial distribution suits as an indicator for the plausibility of our source models. One  disadvantage of previous radial distributions was that usually only one astrophysical object class was taken into account.  Exemplary for an early study of this kind was \citep{case_new_1998}  in 1998, which used supernova remnants from \citep{Green1996ACO}. Later studies like \citep{yusifov_revisiting_2004} and \citep{lorimer_parkes_2006}, chose pulsars as their main source class. However, all three studies showed one common problem in their Galactocentric radial distributions, which is indicated in figure \ref{fig:radial_dist}. 
They all steeply decrease towards the Galactic center and some even revised their distributions upwards to reach source densities different from zero in the Galactic Center. This made them more plausible as there is evidence for sources in this region, not least the Galactic center source itself \citep{Anjos_2020}.  A more recent study of the Galactocentric distribution of sources was performed in \citep{Linden2016}. Motivated from the apparent deficiency in $\gamma$-ray flux around the Galactic center they corrected for the apparent lack of sources in previous studies by adapting the source distribution via including a $20~\%$ fraction of CRs, injected in predicted star-formation regions traced by molecular H$_2$, for which they used the model by \citep{Pohl_2008}.The combination of sources, in this case SNRs, as used in \citep{case_new_1998}, and CR sources tracing the molecular gas density is strongly preferred in comparison to the previous models, since it shows an increase in the Galactic center region, as shown in figure \ref{fig:radial_dist}.
The radial distributions stemming from our source models all show lower values around the Galactic center than the distribution by \citep{Linden2016}. This lack of sources could be linked to the absence of a central Galactic bar in the underlying simulated source sample \citep{steppa_constantin_model_nodate}. At larger radii the radial distribution of our model is in qualitative agreement with the previous models according to figure \ref{fig:radial_dist}.

\section{CR transport simulation}\label{sec:sim_setup}
The source samples from \textit{models 1} - \textit{4} are implemented for use in the PICARD code, which is built around the theoretical understanding of CR propagation in the interstellar medium and computes a steady-state solution of the CR transport equation for a given CR source distribution. For a detailed description of PICARD we refer to \citep{kissmann_picard_2014}.
The modelled processes in the ISM include for example spatial diffusion and diffusive acceleration as well as nuclear spallation, secondary particle production, and different energy loss processes. These energy losses are ionisation and Coulomb losses for all particles and bremsstrahlung, inverse-Compton, and synchrotron losses for leptons only.
Therefore, since leptons and hadrons interact differently with the interstellar medium during propagation due to their large mass difference, every source needs to be defined either as leptonic, hadronic or composite to be included in the transport simulation.
For the firmly identified HGPS-sources with known distances we adopt the source classes as shown in \citep{hess_collaboration_hess_2018}, and assume PWNe and gamma-ray binaries to be leptonic sources, SNRs to be hadronic sources and composites to be sources for both leptons and hadrons. For the part of the synthetic sources taken from the simulation \citep{steppa_constantin_model_nodate} we assume that the fractional division into leptonic, hadronic and composite sources is the same as among the firmly identified HGPS sources with known distances and randomly split the simulated sources accordingly. In particular, we use a percentage of $48\%$ leptonic sources, $26\%$ hadronic sources and $26\%$  sources producing both sorts of CRs. We consider the blend of parent particle distributions from the HGPS representative for the whole Galaxy. Furthermore, a calculation of extreme models, considering all simulated sources accelerating only leptons, hadrons or both, shows that the accelerated particle species only has negligible influence on the outcome of the CR transport simulation as such variations predominantly alter the CR propagation in distant parts of the Galaxy and is, thus, not observable at Earth. A higher fraction of leptonic sources would, however, have an impact on the diffuse flux from unresolved sources.
\subsection{CR transport model setup}
For modelling the CR transport in the Galaxy, a Cartesian spatial grid with $257$ x $257$ x $65$ grid points for the $x$-,$y$- and $z$-axis, respectively,  is used. This spans a domain of $-20~$kpc to $20~$kpc in the $x$-and $y$-direction and $\pm 4~$kpc in the $z$ direction with the Solar System on the positive $x$-axis, employing the IAU-recommended distance from the Sun to the Galactic center $R=8.5$\,kpc \citep{10.1093/mnras/221.4.1023}. This leads to a spatial resolution for the $x$- and $y$ dimension of $0.1556~$kpc and for the z-dimension of $0.123~$kpc. The simulations are carried out covering an energy range from $10~$MeV to $1~$PeV using $127$ logarithmically equidistant points in momentum.  The CR sources are included in the code as 3D Gaussians using the same radius for each source, namely $0.075$\,kpc. A list of the relevant transport parameters we used for all our models is given in table \ref{tab:transport_parameters}. The diffusion coefficient index $a$ is considered a free parameter to reproduce data and can in principle be any value in the range of around $a \approx 0.3-0.6$ \cite{DIBERNARDO2010274}, in this case we use $a=0.31$ \cite{Trotta_2011}. The influence of different halo heights $z_h$ is also investigated in  \cite{Trotta_2011}, where we round their result of $z_h=3.9$\,kpc to $z_h=4$\,kpc. Changing the halo height of the Galaxy would naturally have an impact on the results, as it was investigated in \citep{Kissmann_2015}. They found that a larger halo height results in larger values for the diffusion coefficient, which in turn means that the resulting distributions are more smeared out. Hence, even though we did not investigate the impact of changing the halo height here, the general effect was analysed in \citep{Kissmann_2015}.  Furthermore, the electron injection spectrum values, namely spectral indices and breaks, as listed in table \ref{tab:transport_parameters} are taken from \cite{Orusa_2021} and \citep{refId0}. The nuclei injection spectrum parameters listed in table \ref{tab:transport_parameters} are taken from \citep{2020ApJ...889..167B}, where we used the "propagation"-scenario. This scenario assumes that the break at  400 GeV/nucleon results from a change in the spectrum of interstellar turbulence.
\begin{table}
\caption{Transport parameters used for all models in this study.}
\label{tab:transport_parameters}
\begin{center}
\begin{tabular}{ll} \toprule
 Parameter &  \\
\midrule
\textit{ General}&\\
Halo height  & $4$\,kpc \\
Galactic radius &  $20$\,kpc\\
Diffusion coefficient $D_{xx}~^1$  & 6.3 $\cdot 10^{28}$\,m$^2$/s\\
Diffusion coefficient index  $a$ & $0.31$\\
Alfv\'en speed $v_A$ & $37.04 $\,km/s\\
\textit{ Nuclei injection spectrum} & \\
Index below $1^{\mathrm{st}}$ break &  $1.7$\\
Index between $1^{\mathrm{st}}$ and $2^{\mathrm{nd}}$ break              &        $2.44$\\
Index above $2^{\mathrm{nd}}$ break &  $2.19$\\
$1^{\mathrm{st}}$  break energy & $6.9$\,GeV \\
$2^{\mathrm{nd}}$ break energy & $400$\,GeV \\
\textit{ Nuclei normalisation} &\\
Normalisation energy& $108$\,GeV\\
Normalisation flux &  $4\cdot10^{-2}$\,m$^{-2}$s$^{-1}$sr$^{-1}$GeV$^{-1}$\\
\textit{ Electron injection spectrum} & \\
Index below $1^{\mathrm{st}}$ break &  $1.4$\\
Index between $1^{\mathrm{st}}$ and $2^{\mathrm{nd}}$ break              &        $2.3$ \\
Index above $2^{\mathrm{nd}}$ break &  $3.5$\\
$1^{\mathrm{st}}$  break energy & $1$\,GeV \\
$2^{\mathrm{nd}}$ break energy & $2.5\cdot10^3$\,GeV \\
\textit{ Electron normalisation} &\\
Normalisation energy& $25$\,GeV\\
Normalisation flux &  $1.21\cdot10^{-2}$\,m$^{-2}$s$^{-1}$sr$^{-1}$GeV$^{-1}$\\
\bottomrule
\\
$^1~~D_{xx}=\beta (R/R_0)^a$ with $R_0=4$GV as reference rigidity
\end{tabular}
\end{center}
\end{table}
Two of those propagation parameters, namely the spatial diffusion coefficient $D_{xx}$ and the Alfv\'en speed $v_A$ have been tuned to reproduce the CR data measured at Earth.  A common test for the reliability of propagation models is the boron to carbon flux ratio (B/C), which we also used to confirm consistency between the PICARD model results and the CR data.
This is because carbon nuclei found in CRs are produced and accelerated in the astrophysical CR sources, whereas boron nuclei are produced exclusively in interactions of heavier CR nuclei with the interstellar matter.
Therefore,  the B/C-ratio estimates the amount of  interstellar material the CR particles passed through on their way to the observer and can, hence,  be used to constrain the spatial diffusion coefficient \citep{B/C_by_AMS}.  

\begin{figure}
	\centering
		\includegraphics[scale=.5]{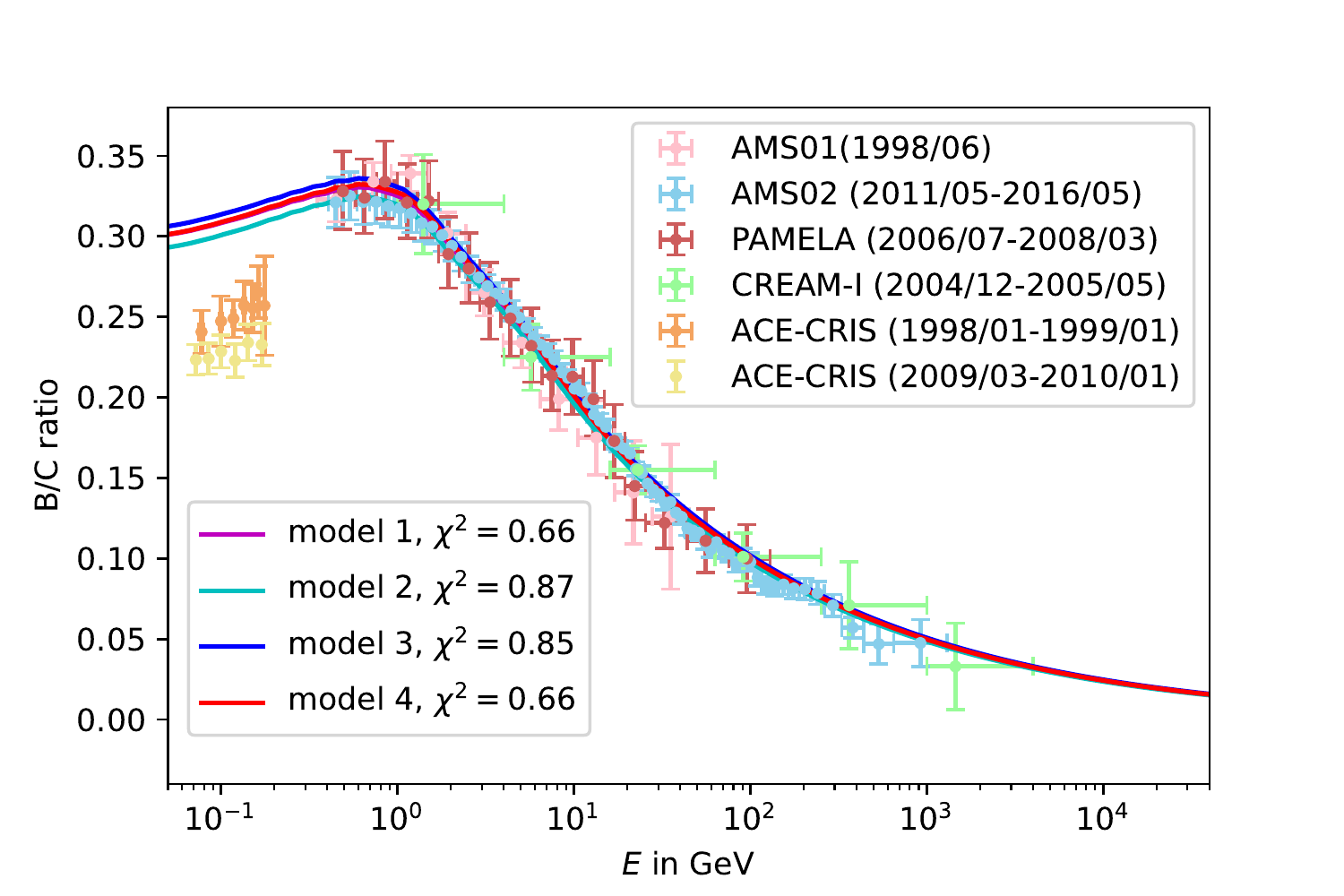}
	\caption{B/C-ratio for our \textit{models 1-4} in comparison with CR data taken from different experiments (top to bottom in the legend:\citep{aguilar_relative_2010}, \citep{aguilar_precision_2016}, \citep{adriani_measurement_2014}, \citep{ahn_measurements_2008}, \citep{de_nolfo_observations_2006}, \citep{lave_galactic_2013}) and the corresponding values for the goodness of fit.}
	\label{fig:B/C_ratio}
\end{figure}

The B/C-ratio for our models in comparison with CR data is shown in figure \ref{fig:B/C_ratio}. 
The best fit parameters were optimized such that all models deliver good results for the secondary to primary ratios. We found that the diffusion coefficient $D_{xx}=6.3\cdot 10^{28}~$m/s and the Alfv\'en speed $v_A=37.04~$km/s lead to an agreement with observations by different instruments and the simulations, especially in the high-energy regime, which is most relevant for our study of very-high-energy diffuse $\gamma$-ray emission.
Furthermore, at lower energies solar modulation influences the observed data, which depends highly on the applied heliospheric modulation model, where we use a simple force field approximation.  In addition, both observations by ACE-CRIS were conducted during solar minimum periods  and have, therefore, not been taken into account for the calculation of the $\chi^2$ but are shown in the plot for completeness. Quantitatively the goodness of fit test of the simulated B/C-ratios in comparison to observations lies between $\chi^2=0.66$ and $\chi^2=0.87$. 
Since we want to investigate the impact and the resulting differences of the underlying source distributions we do not vary the propagation parameters between the models, even though this would improve the fit. 
In contrast to our values of $D_{xx}=6.3\cdot 10^{28}~$m/s and $v_A=37.04~$km/s for the diffusion coefficient and the Alfv\'en speed.

\section{Results}\label{sec:result}
In this section we present results of our transport models using the four different source distributions by starting to discuss the spatial distributions of CR electrons and protons. We compare the particle spectra at Earth with measured data and analyse particle spectra at different positions in the Galaxy. Finally, we show predicted $\gamma$-ray fluxes, calculated with the new source distributions and the respective $\gamma$-ray spectra in different directions in the Galactic plane.  We conclude by quantifying the influence of the difference between our models on the propagation results.
\subsection{CR density distributions}\label{ssec:CR_dens_dist}
\begin{figure*}
	\centering
		\includegraphics[width=\textwidth]{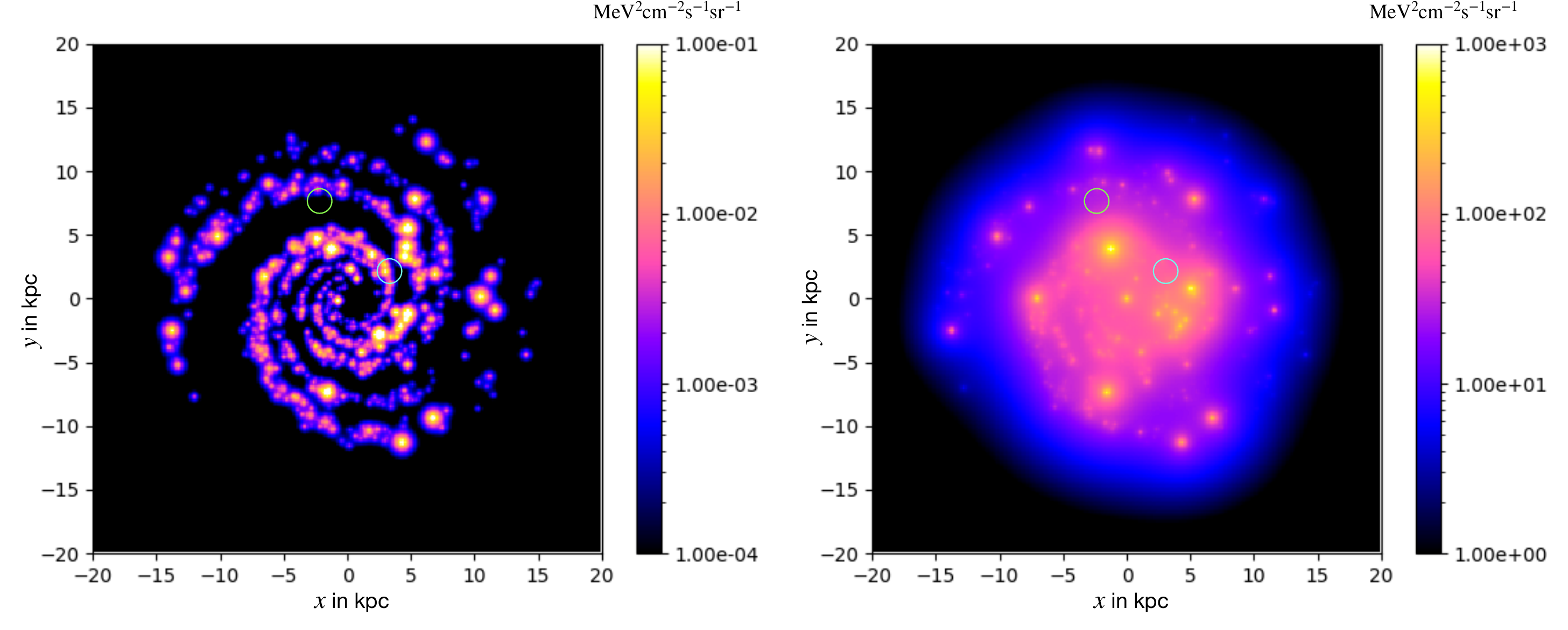}
	\caption{\textit{Left}: Density distribution of simulated CR electrons with an energy of $10.76~$TeV in the $x$/$y$-plane of the Galaxy using the HGPS sources and the sources from \textit{model 1}, exemplarily, shown with a logarithmic scale for better visibility. In this representation the Solar System location is assumed to be at $(8.5,0)~$kpc. The light blue circle represents the region within the Norma Arm and the light green circle the interarm region, specifically discussed in section \ref{ssec:CR_spec}. \textit{Right}: Same but for protons.}
	\label{fig:CR_density}
\end{figure*}

In figure \ref{fig:CR_density} we show slices through the electron and the proton distributions in the Galactic plane, exemplary for \textit{model 1} at an energy of $10.76~$TeV.
Particularly, in the electron distribution (left) the underlying source model is visible and the sources observed in the HGPS are showing very prominently, like for example the PWNe HESS J1303-631, HESS J1418-609, HESS J1420-607, and HESS J1514-591 aligned at $x\approx 5$\,kpc (ranging from $y=4$\,kpc to $y=5.5$\,kpc). This very clear imprint of the sources results from the severe energy losses of high-energy electrons, allowing $10$\,TeV electrons to propagate only a small distance before they loose most of their energy. 
For protons (right) the distribution is more extended given their much larger energy-loss scale, although the imprint of the underlying source model is still visible and we also see several very luminous sources stemming from the HGPS observations, like for example the SNR HESS J1718-374 at $(-3.1,2.1)$\,kpc.
\subsection{CR spectra}\label{ssec:CR_spec}
\begin{figure*}
	\centering
		\includegraphics[width=\textwidth]{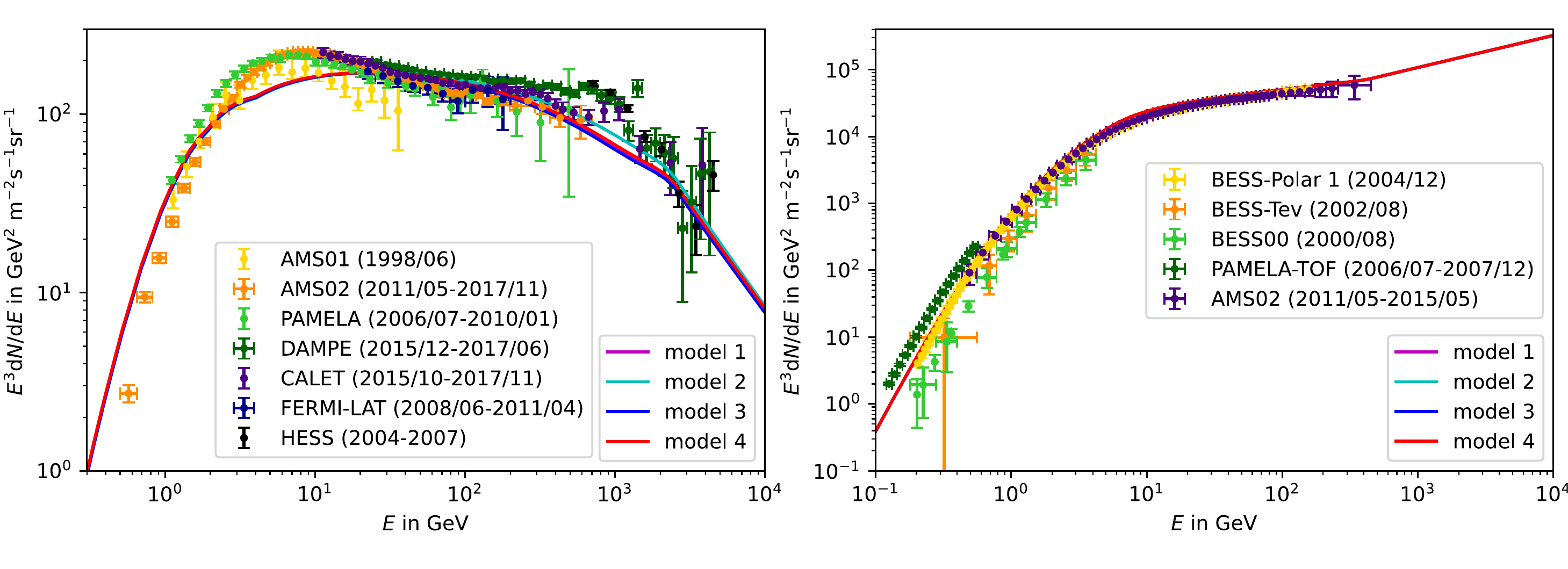}
	\caption{\textit{Left}: Electron spectra at for the four different models in comparison with each other and with observed data from different experiments (top to bottom in the legend: \citep{aguilar_relative_2010}, \citep{ackermann_measurement_2012}, \citep{adriani_cosmic-ray_2011},\citep{Aharonian_2008}). We applied a force field parameter of $\Phi=647\,$MV during the propagation, which is the force fields found in \citep{ackermann_measurement_2012}. \textit{Right}: Same but for CR protons, with a force field parameter of  $\Phi=685\,$MV, which is the eference value from \citep{adriani_measurements_2016}. Experiments from top to bottom: \citep{abe_measurement_2008}, \citep{shikaze_measurements_2007}, \citep{shikaze_measurements_2007}, \citep{adriani_measurements_2016}.}
	\label{fig:ep_spectra_earth}
\end{figure*}

The proton and electron spectral intensities at Earth are shown in figure \ref{fig:ep_spectra_earth}, where we compare \textit{models 1-4}, obtained in this study, with CR observations. 
To quantify the difference between the four models we integrate over each spectrum from $0.1$\,GeV to $10$\,TeV, which leads to an average standard deviation between the four spectra, of $4.2\%$ for the electron spectra and $0.1\%$ for the proton spectra.
This small difference for spectra at Earth is expected because the proton and the electron spectra are normalised in our simulations. This means that the primary electron fluxes are scaled with a constant factor so that their flux at the normalisation energy given in table \ref{tab:transport_parameters} is identical to the given normalisation flux. The same is done with protons, where all nuclei and also all secondaries, are scaled with the same factor as the protons. Because, apart from that, the transport parameters (here we only adapted the spatial diffusion coefficient and the Alfv\'en speed) are modified in a way, that the observed B/C ratio at Earth is reproduced with our model, also the fluxes of other nuclei will show no significant variation at Earth.  
Furthermore, we can clearly see the breaks in both particle spectra, which stem from the injection spectra used for the simulation as mentioned above and shown in table \ref{tab:transport_parameters}.

\begin{figure*}
	\centering
		\includegraphics[width=\textwidth]{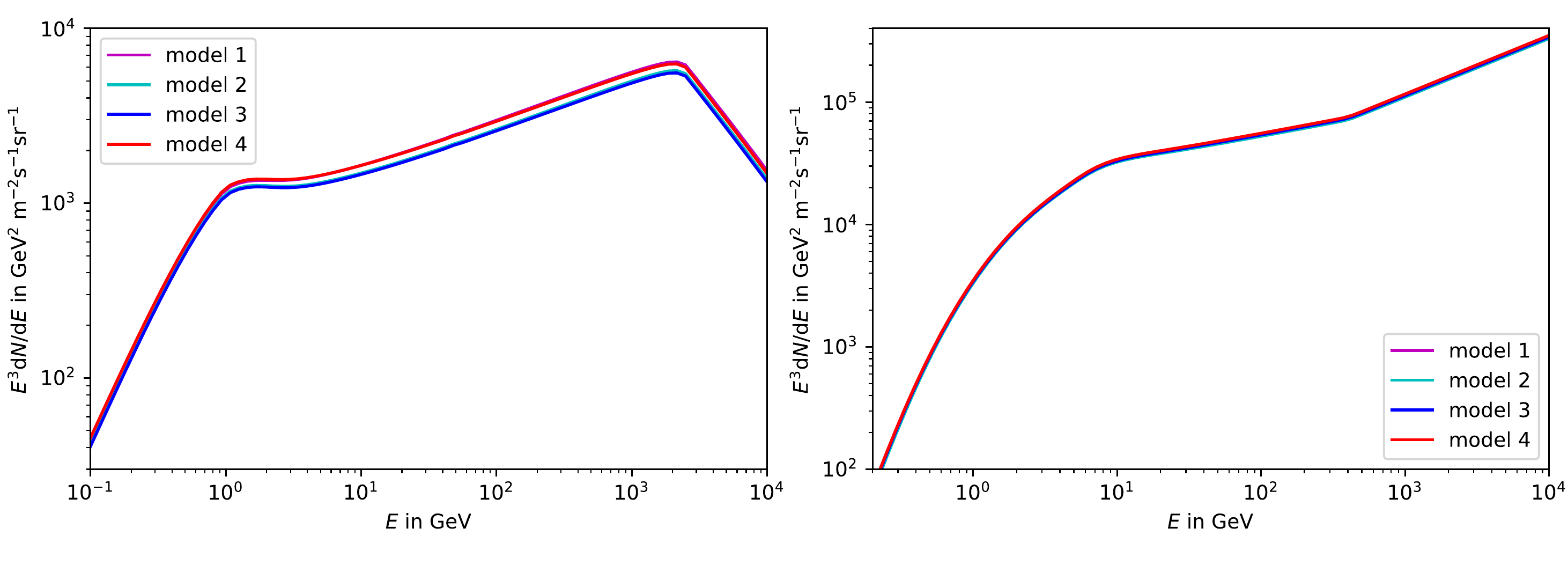}
	\caption{\textit{Left}: Electron spectra at a position of $(3,2)$\,kpc within the Norma Arm for the four different models in comparison with each other. \textit{Right} Same but for CR protons. }
	\label{fig:spectra_arm}
\end{figure*}

To avoid this normalisation-induced bias we also study the particle spectra at other positions in the Galaxy.  To illustrate deviations between our \textit{models 1-4} we analyse the spectra at two specific points in the Galactic Plane, one within a spiral arm, in this case the Norma Arm, at $(3,2)$\,kpc and one in an interarm region at $(-2,7.2)$\,kpc, which are shown in figure \ref{fig:CR_density} as light blue and green circles, respectively. In figure \ref{fig:spectra_arm} we show the electron and proton spectra within the Norma arm for all four models.
In comparison to the spectra at Earth we note qualitatively larger deviations in the electron spectra (left) at the Norma arm,  because here we do not have local normalisation but larger changes concerning the source density and distribution. The proton spectra (right) at the Norma arm qualitatively look quite similar. We calculate the average deviation of the integrated spectra from $0.1$\,GeV to $10$\,TeV to quantify deviations between the different models. From the integral fluxes within the spiral arm region we obtain an average spread of $6\%$ for the electron spectra and $4\%$ for the proton spectra. Correspondingly, the difference increased by almost $4$\% in comparison to the spectra at Earth whereas for the electrons it increased about $2$\%. 
\begin{figure*}
	\centering
		\includegraphics[width=\textwidth]{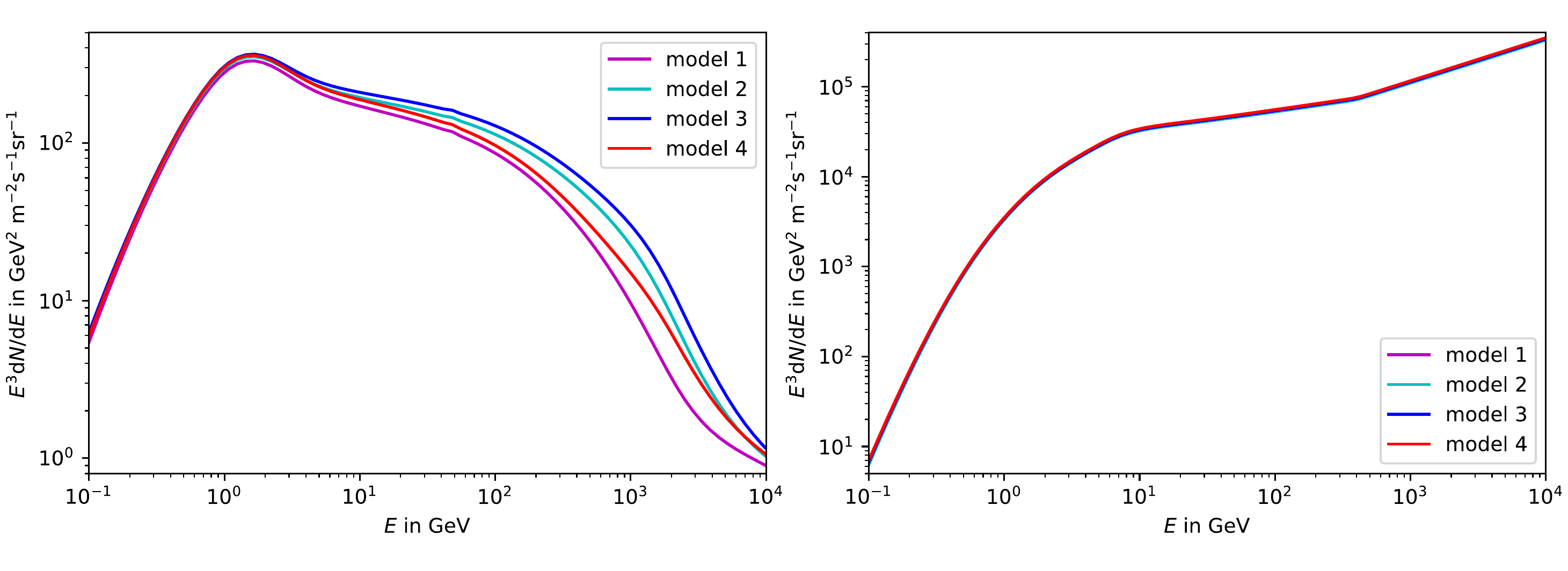}
	\caption{Same as \ref{fig:spectra_arm} but for the interarm region at $(-2,7.2)$\,kpc.}
	\label{fig:spectra_interarm}
\end{figure*}

In figure \ref{fig:spectra_interarm} we show the spectra in the interarm region, which show larger differences between our \textit{models 1-4} in the high-energy regime of the electron spectrum, whereas the proton spectra and also the low-energy part of the electron spectrum are very similar in this region. This is because protons and low-energy electrons are not strongly affected by energy-loss processes and will, therefore, not be able to reach high distances from the source. This leads to the relative deviation of the integral over the interarm region spectrum giving $24\%$ for the electron spectra and $2\%$ for the proton spectra.  Especially for the electron spectra this still seems smaller than the qualitatively visible difference. This similarity despite the obvious differences stems from the small influence of the high energy part of the spectrum to the integral. If the integration is carried out from $100$\,GeV to $10$\,TeV instead the average deviation for the electron spectra would be $32\%$, whereas for the proton spectra it would make no difference. 

\subsection{Gamma-ray fluxes}\label{ssec:gamma_map}
\begin{figure*}
	\centering
		\includegraphics[width=\textwidth]{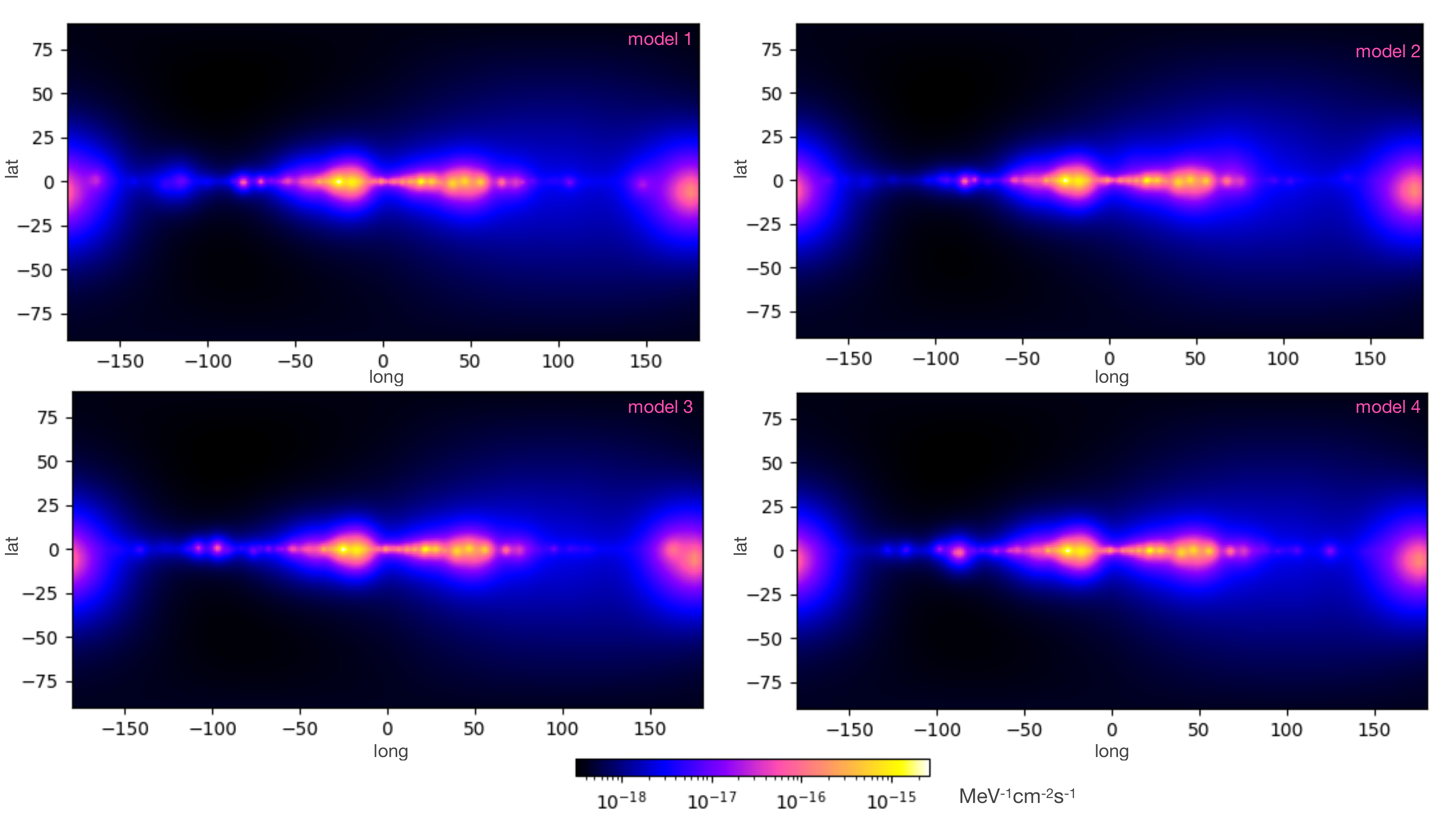}
	\caption{Inverse Compton flux at an energy of $1$\,TeV, for \textit{models 1-4}. }
	\label{fig:gamma_map_ic}
\end{figure*}

The calculations for the $\gamma$-ray emissivity and the corresponding $\gamma$-ray flux are done separately by line-of-sight integration for the relevant radiation processes, namely pion decay, inverse Compton scattering, and bremsstrahlung.  In figure \ref{fig:gamma_map_ic} we show the fluxes of inverse Compton scattering for our \textit{models 1-4} at an energy of $1$\,TeV for the whole Galaxy.
We chose to depict the inverse Compton skymaps because the high-energy electrons only occur close to their sources due to their high energy losses and, therefore, clearly show the different sources simulated within each model. This leads to an increase in $\gamma$-ray fluxes near the spiral arm tangents of the underlying synthetic source population.  In figure \ref{fig:gamma_map_ic} we can see several differences between our \textit{models 1-4}, stemming from sources being present in one model but lacking in the next.  The most obvious differences are visible around $100^{\circ}$ in the Galactic plane, stemming from the tangent of the tail of the Cygnus arm.
\begin{figure*}
	\centering
		\includegraphics[width=\textwidth]{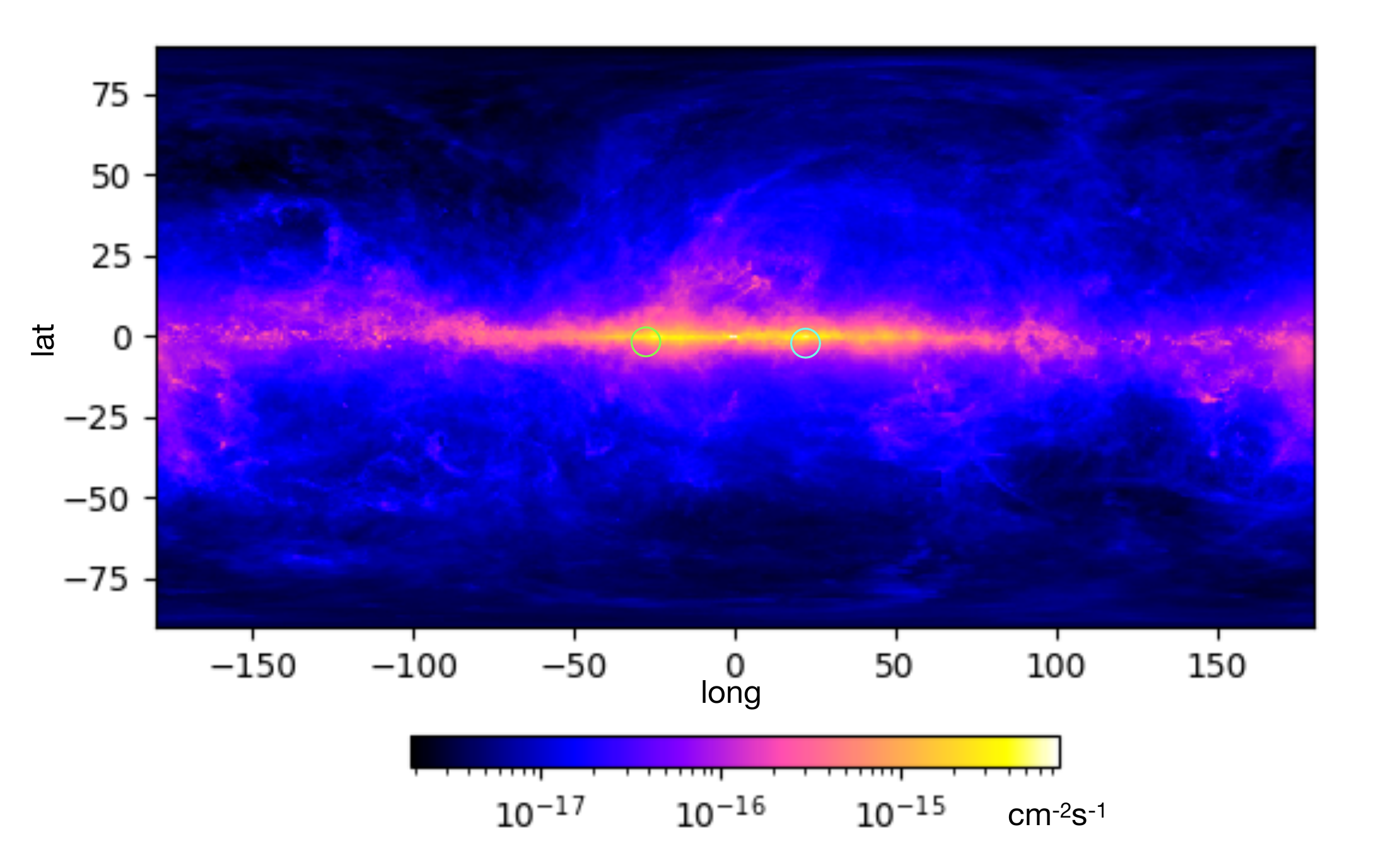}
	\caption{Total $\gamma$-ray flux at an energy of $1$\,TeV (\textit{model 1} only). The light green circle represents the region within the Norma Arm and the light blue circle the interarm region, specifically discussed in section \ref{ssec:gamma_spec}.}
	\label{fig:gamma_map_tot}
\end{figure*}

In figure \ref{fig:gamma_map_tot} we depict the flux of the total $\gamma$-ray flux for the whole Galaxy, exemplarily for our \textit{model 1} at an energy of $1$\,TeV. In comparison to the inverse Compton scattering the total $\gamma$-ray flux projection shows only a weak imprint of the spiral arms because Galactic diffuse $\gamma$-ray emission is dominated by pion decay. 
In pion decay, as well as bremsstrahlung, the emissivity follows from a convolution of the interstellar gas distribution with the cosmic-ray fluxes. This results in only minor differences between our four models in the total $\gamma$-ray flux.

\begin{figure*}
	\centering
		\includegraphics[width=\textwidth]{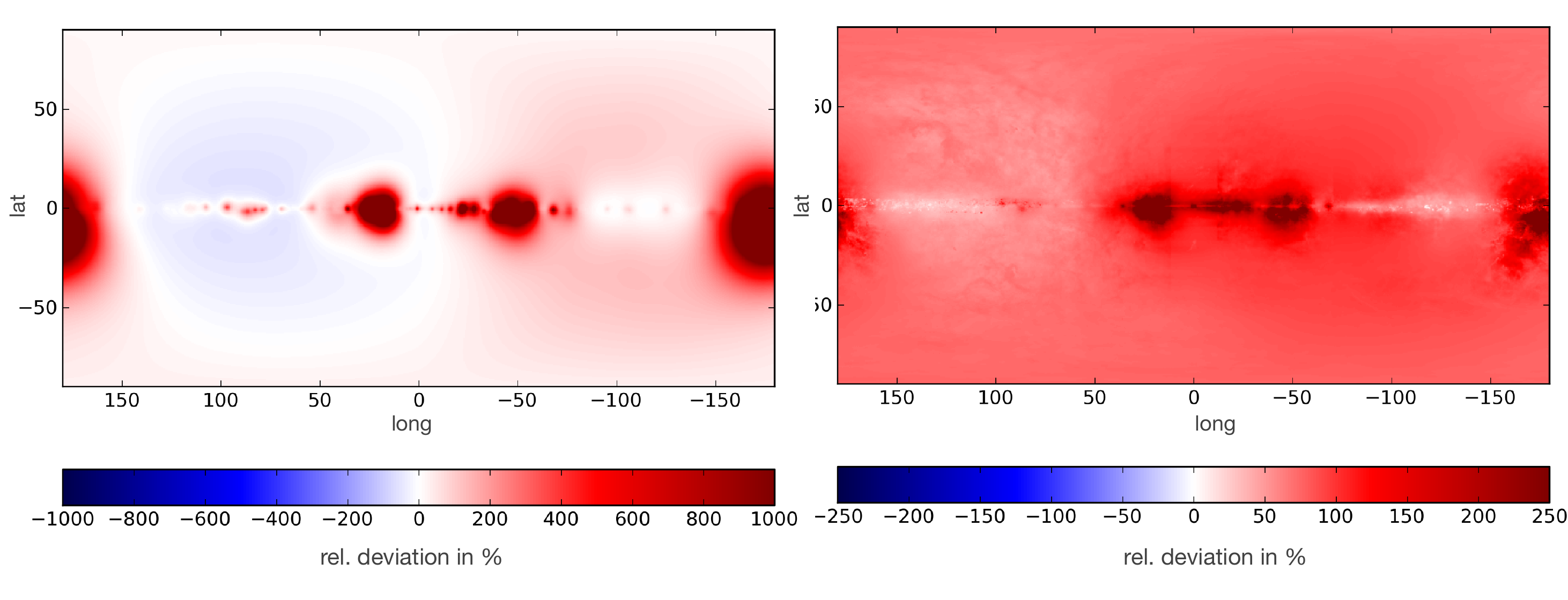}
	\caption{\textit{Left:} Relative deviations of our \textit{models 1-4} in comparison to the axially symmetric model \citep{Kissmann_2015}, for the inverse Compton scattering $\gamma$-ray flux and covering the whole Galaxy at an energy of $1$\,TeV  \textit{Right:} Same for the total $\gamma$-ray flux.}
	\label{fig:gamma_map_std_dev_vgl}
\end{figure*}

Ours results can be qualitatively compared to axially symmetric analytical $\gamma$-ray sky models from literature, e.g.  as used in \citep{Kissmann_2015}, which is shown in figure \ref{fig:gamma_map_std_dev_vgl}.
Thereby, we plot the relative deviation between the mean value of our \textit{models 1-4} and the previous axisymmetric analytical model for both inverse Compton scattering as well as total $\gamma$-ray emission in figure \ref{fig:gamma_map_std_dev_vgl}. For the total emission we find an average deviation of $83$\%. Local variations go up to $831$\%  and are visible at longitudes between $30^{\circ}$ and $-50^{\circ}$ in the Galactic plane, which coincides with the location of the Norma arm. The position within a spiral explains the difference between our models and the previous model since we add sources specifically according to the simulated sample from \citep{steppa_constantin_model_nodate}, which follows the matter density in the Galaxy. The relative deviations between our model and the axially symmetrical model are larger for the inverse Compton scattering channel, namely at an average of $107$\% with local fluctuations up to $-58$\% and $10663$\%, which are found within the regions of the Centaurus and Norma arms, at longitudes of about $-50^{\circ}$ and $20^{\circ}$ in the Galactic plane, respectively.

\begin{figure*}
	\centering
		\includegraphics[width=\textwidth]{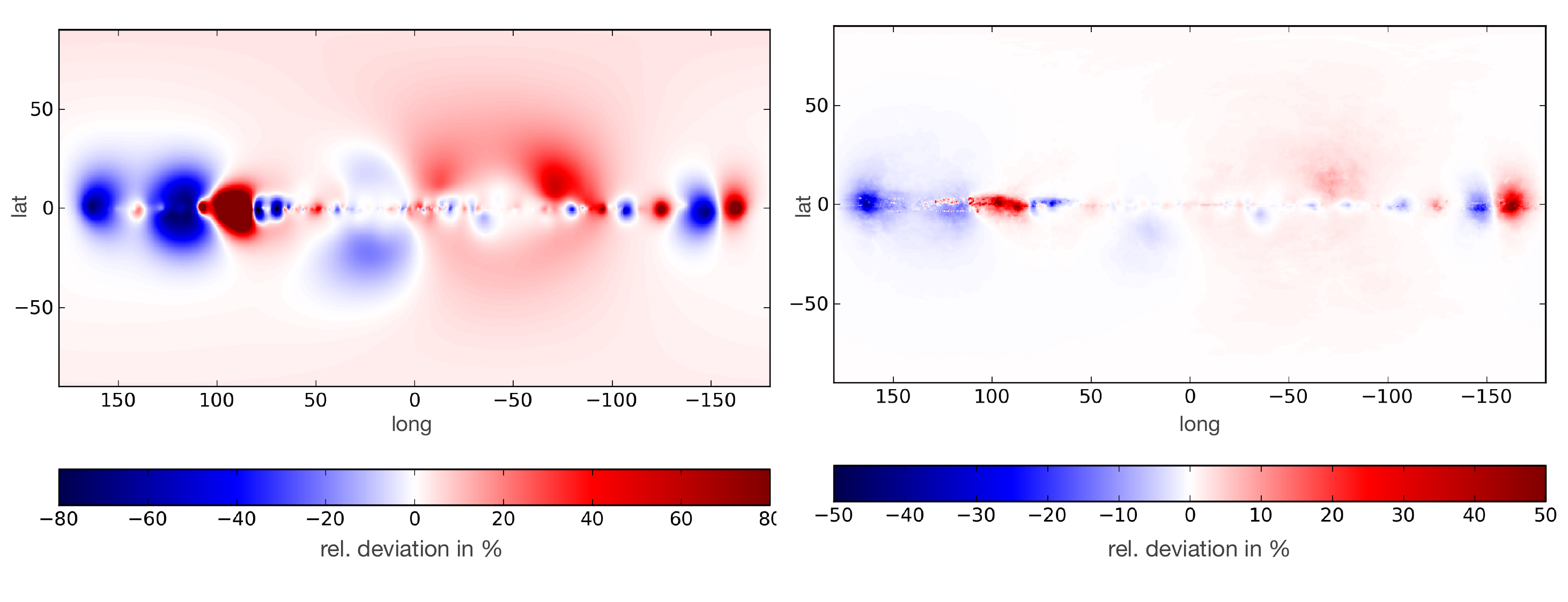}
	\caption{\textit{Left:} Relative deviations of our \textit{model 1} and \textit{models 2-4} for the inverse Compton flux for the whole Galaxy at an energy of $1$\,TeV  \textit{Right:} Same for the total $\gamma$-ray flux. }
	\label{fig:gamma_map_std_dev}
\end{figure*}

For another quantitative interpretation of the $\gamma$-ray emission in our new source models we also calculate the average deviation between our \textit{model1} and \textit{models 2-4}, showing it in figure \ref{fig:gamma_map_std_dev}.
Thereby, we find that the global average deviation between our \textit{model 1} and \textit{models 2-4} for the total $\gamma$-ray calculations lies at $0.27$\% whereas locally we find deviations up to $-41$\% and $91$\%, e.g. visible in the Galactic plane at longitudes of about $100^{\circ}$ on the right side of figure \ref{fig:gamma_map_std_dev}, which corresponds to the position of the tail of the Cygnus arm. For the inverse Compton $\gamma$-ray channel, as expected, the differences between the individual models are higher, namely showing an average deviation of $3.7$\% with individual peaks of up to $-87$\% and $1132$\%. The highest deviations occur near the spiral arm tangents, where we also have the largest number of sources and correspondingly the largest differences between the four models.
Such signatures are visible very prominently, e.g., in the Galactic plane at longitude $90^{\circ}$ in the left part of figure \ref{fig:gamma_map_std_dev}, which is also related to the tail of the Cygnus arm. 

\subsection{Gamma-ray spectra}\label{ssec:gamma_spec}
\begin{figure}
	\centering
		\includegraphics[scale=.55]{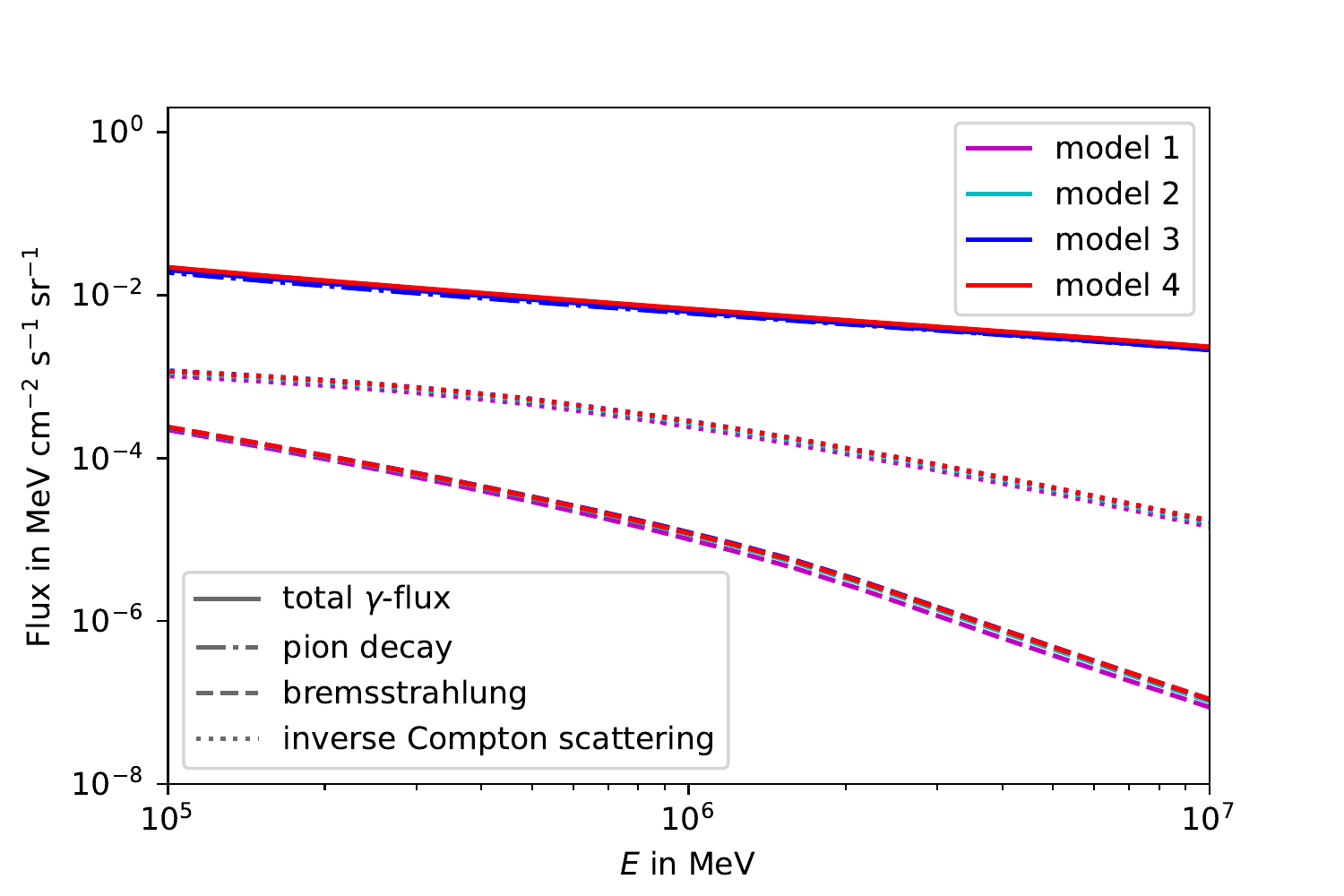}
	\caption{$\gamma$-spectra for pion decay, bremsstrahlung and inverse Compton scattering and the total $\gamma$-ray flux, simulated for our \textit{models 1-4} in the direction of the Galactic Center $l=0^{\circ}\pm1^{\circ}$.}
	\label{fig:gamma_spec_earth}
\end{figure}

After analysing the $\gamma$-ray flux at an energy of $1$\,TeV we now investigate the underlying spectral shape. We calculate the $\gamma$-ray spectra for each of the processes mentioned above for three different, specifically selected directions in the Galactic Plane, namely in the direction of the Galactic Center, of the spiral arm tangent of the Norma Arm, and of an interarm region. Results for the Galactic Center are shown in figure \ref{fig:gamma_spec_earth} for all our models. 
We see that all four of our models give similar overall results, as well as in each individual radiation process, and are constrained by the normalisation to the CR fluxes at Earth and the flux attributed to the unresolved sources from \citep{steppa_constantin_model_nodate}. We quantify differences between our models by comparing the relative deviations between our \textit{models 1-4} for each spectrum. This is done in the same way as for the particle spectra by integrating over each spectrum from $0.1$\,TeV to $10$\,TeV.  Therefrom, we find relative deviations, as listed in table \ref{tab:gamma_spec}, for the different emission processes.
\begin{table}
\caption{Average relative deviations (in \%) between our \textit{models 1-4} for bremsstrahlung (B), inverse Compton scattering (IC), pion decay (PD), and the total gamma-ray flux (TOT) for the three different investigated directions in the sky: the Galactic Center, the Norma arm tangent at $l=25^{\circ}\pm1^{\circ}$ and an interarm region at $l=-33^{\circ}\pm1^{\circ}$.}
\label{tab:gamma_spec}
\begin{center}
\begin{tabular}{lcccc} \toprule
  & B & IC & PD & TOT\\  \midrule
Galactic Center & $3.8$ & $5.8$ &$ 3.2$ & $3.1$ \\
Norma arm region & $2.6$ & $2.5$& $2.9$ & $2.7$ \\
Interarm region & $2.8$ &$ 4.5$ & $2.2$ & $2.3$ \\
 \bottomrule
\end{tabular}
\end{center}
\end{table}
Changes of the source distributions are small for most interaction channels near the Galactic center, since we do not have a Galactic bar present in the underlying simulated source model, rendering all models similar in the Galactic center, namely lacking simulated sources. However, inverse Compton scattering is showing a higher deviation due to the longer interaction lengths. 

\begin{figure}
	\centering
		\includegraphics[scale=.55]{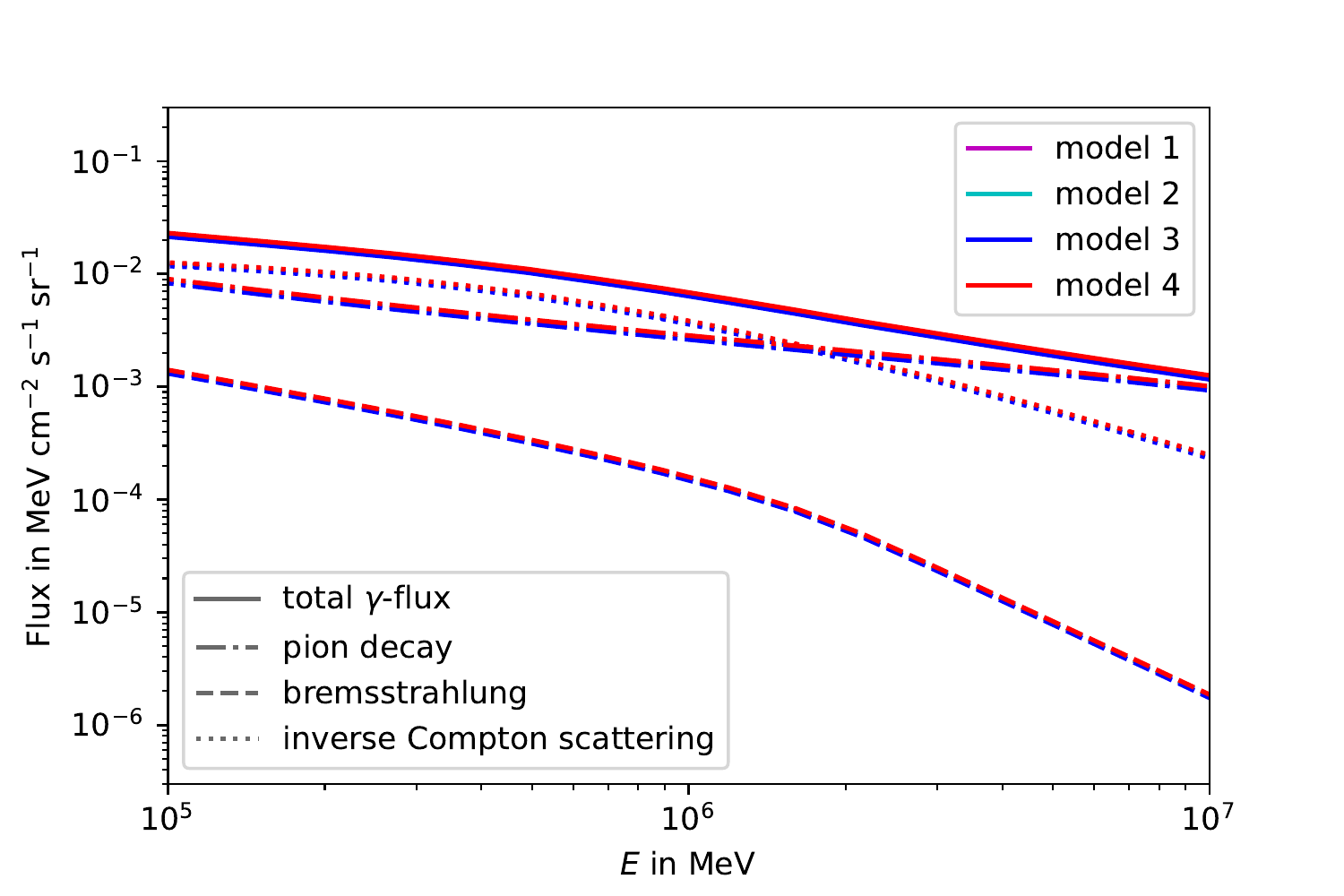}
	\caption{Same as figure \ref{fig:gamma_spec_earth} but in the direction of the Norma arm tangent at $l=25^{\circ}\pm1^{\circ}$.}
	\label{fig:gamma_spec_arm}
\end{figure}

In figure \ref{fig:gamma_spec_arm} we show the results in the direction of the Norma arm tangent at $l=25^{\circ}\pm1^{\circ}$. 
This point was chosen to investigate the largest local differences since the source distributions are most different in the spiral arms.  However, the deviations between our models are still low, in the same order as in the direction of the Galactic center.
One notable feature in the spectra of figure \ref{fig:gamma_spec_arm} is that inverse Compton scattering is having a considerably higher influence on the overall $\gamma$-ray flux, in the lower energy regime even exceeding the pion decay flux.
The relative deviations in the direction of the Norma arm tangent, reported in table \ref{tab:gamma_spec}, all show similar low values.
While in the direction of the Galactic center as well as the interarm region both the deviations of pion decay and of the total flux are similar, since the total flux is dominated by pion-decay emission, in the direction of the Norma arm we see a larger deviation between pion decay and the total $\gamma$-ray flux.
This increase in inverse-Compton gamma-ray emission in the direction of a spiral arm stems from the short interaction wavelengths of leptonic particles, which peaks in the region which contains most sources.

\begin{figure}
	\centering
		\includegraphics[scale=.55]{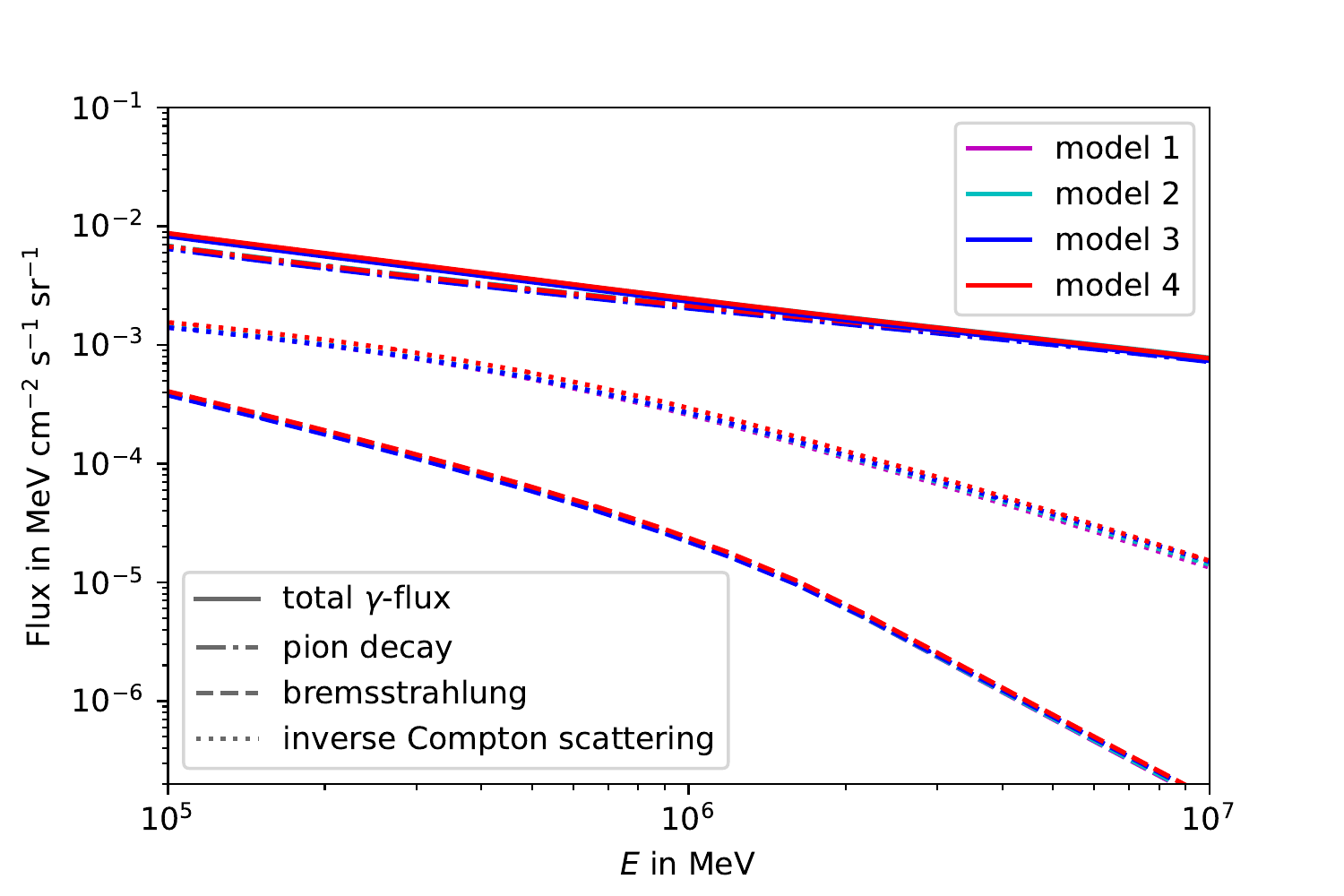}
	\caption{Same as figure \ref{fig:gamma_spec_earth} but in the direction of an interarm region at $l=-33^{\circ}\pm1^{\circ}$.}
	\label{fig:gamma_spec_interarm}
\end{figure}

The result for the interarm region in the direction of $l=-33^{\circ}$, is shown in figure \ref{fig:gamma_spec_interarm}. 
Here, we also find similar deviations (see table \ref{tab:gamma_spec}) between the models in most channels, given that in this direction there are no sources in any of the models, which also explains the higher deviations in the inverse Compton channel. This is also visible in the pion decay spectrum again being the most dominant emission channel by one order of magnitude.

\section{Discussion}\label{sec:disc}
Our hybrid cosmic-ray source model is constrained by its two principal inputs: the observed and the simulated source sample. This combination offers several benefits, namely combining the accuracy of the observed sources with the multitude of the simulated sample. One issue we face thereby is that potential disadvantages of the synthetic source model will be passed on to our models like the absence of a Galactic bar in the underlying source model. As a result our models are missing sources in the Galactic center region. 
While previous Galactocentric source distributions (\citep{case_new_1998}, \citep{yusifov_revisiting_2004}, \citep{lorimer_parkes_2006}, \citep{Linden2016}) were essentially based on radial symmetry and analytical parametrisations, we now use the full information of our three dimensional source distribution.
Our distribution agrees well with the previous analytical distributions for high Galactocentric radii but underestimates the source amount in the Galctic center region.

In our simulations of CR propagation with PICARD, the imprint of the underlying model for the CR-source distribution is evident in the CR density distributions of electrons and protons, distinctly showing the location of sources in the spiral arms. At particle energies above $1$\,TeV we can see clear differences between the two particle species, as anticipated from their energy loss lengths. 

Simulated particle spectra at Earth offer the possibility to compare our models with actual experimental data. Generally, they show good agreement, which is also not surprising due to the normalisation of the code and because the models were tuned to fit the B/C-ratio data. Apart from that we also find our models in reasonable consistency with each other.
Furthermore, we are also interested in localised deviations between the models and evaluate them at different, specifically motivated regions in the Galaxy. We chose two locations, one within the Norma arm and one in an interarm region. Spectra simulated within the spiral arm show larger deviations between the \textit{models 1-4} because the source composition changes significantly more than in the interarm region, where our models do not have sources and instead we only observe primary CRs transported from their sources in the spiral arms and signals of locally produced secondaries. 
Therefore, in the interarm region secondary electrons play a major role which is reflected in the larger differences between our models in the electron spectra, whereas the proton spectra are similar for all models at all three different locations because they are manly primaries.

We note that the individual sources are not showing in the total $\gamma$-ray flux of our \textit{models 1-4} because it is dominated by pion-decay emission, where the emissivity follows from a convolution of the interstellar gas distribution with the cosmic-ray fluxes. As a result, we can only see minor differences between the total $\gamma$-ray emissions of our \textit{models 1-4}.  For the flux of inverse Compton scattering the differences are higher and peak in the directions of the spiral arm tangents. 
This is also visible in the comparison of our models with a previous models using an axisymmetric CR-source distribution. Since we do not have sufficient information about the Galaxy to create a singular perfect model of the CR source-distribution
  
We also analysed the $\gamma$-ray spectra, for each interaction channel and each model. For that, we choose three specifically motivated directions, namely the direction towards the Galactic Center, towards the Norma arm tangent and towards an interarm region. An interesting aspect found there is that the inverse Compton scattering component of the $\gamma$-ray flux is dominating in the direction of the Norma arm.

We find that our \textit{models 1-4} all yield globally consistent results for CR propagation and that the simulated source number in each model is rendered negligible, which is a benefit of using the same underlying simulated source model. Since we do not have sufficient information on the Galaxy and its composition it is not yet possible to create a singular perfect of the CR source distribution but one should work with a statistical sample of possible models.

When comparing our results with a recent study assessing the diffuse emission from the unresolved sources \citep{Vecchiotti_2022}, several differences are to note. In contrast to our study they focus on the overall diffuse emission of unresolved sources, in particular PWNe and TeV halos, whereas our focus lies on the CR injection and propagation from individual sources and the ensuing diffuse $\gamma$-ray emission. The truly diffuse flux calculation in  \citep{Vecchiotti_2022} relies on the analytical description by \citep{PhysRevD.98.043003}, whereas this study considers CR propagation in our Galaxy. Lastly, as $\gamma$-rays from PWNe and TeV halos are exclusively related to leptonic emission processes in \citep{Vecchiotti_2022}, our models are dominated by hadronic emission processes, as in the large-scale diffuse flux in \citep{PhysRevD.98.043003}. In order to compare principal results of these studies despite the differences, we plan to explore the boundaries of variations in the flux from unresolved sources by varying the propagation set-up in our models.

While our redefined source model corresponds to an advancement in terms of morphology, all sources as well as the subsequent CR propagation are based on a steady state assumption. A potential further improvement would, therefore, be to apply time-dependent CR injection in the modelling, as discussed in \citep{articlepjm}. Given that most sources, like SNRs and PWNe, are known to be active CR emitters only for a very limited amount of time, a corresponding implementation is expected to yield additional accuracy. However, since the specific history of the Universe and, correspondingly, its general time evolution is uncertain, the approach to implement a more realistic source model leads to less assumptions inherent in the final population. Therefore, we choose to study more realistic source distributions, based on observations before entertaining time dependence.

\section{Conclusion}\label{sec:out}
We present the construction and application of a CR source model of the Milky Way, which goes beyond parametrised axisymmetric distributions. Our model is based on a combination of observed and simulated sources. The observed sources were taken from the H.E.S.S. Galactic Plane Survey, including both those that are firmly identified with known distances and those without any distance estimate, for which we selected a suitable source from the simulated sample. By construction the hybrid model depends on the specific statistical realisation although the source number stays constant within $3$\% for the different models and different realisations do not affect the global simulated CR transport results notably. When comparing our models with those from the literature we find a similarly low source density in the Galactic center region, which in our case relates to the lack of a Galactic bar in the used underlying simulated source sample. Even though a good parametrisation might be superior to a source model, extrapolating the fraction of observed sources to the whole Galaxy allows to estimate the amount of sources currently undetected at TeV energies and point out the expectations for forthcoming observations with future very high-energy $\gamma$-ray experiments.

We are able to reproduce measured secondary to primary ratios as well as CR particle spectra measured at Earth. Considering that our simulation is normalised to the position of the Earth, we analyse electron and proton spectra at the Norma arm and an interarm region to represent a more thorough image of the Galaxy. Even though we expect larger contrast between the models in those regions, we find consistency between the particle spectra of the different model realisations in the spiral arm while in the interarm region the contribution of secondary electrons increases the deviation between the spectra.

We calculate $\gamma$-ray flux and energy spectra with PICARD, in the different relevant interaction channels.  We present $\gamma$-ray flux predictions and spectra in different directions in the Galaxy, i.e. the direction of the Galactic Center and, again, of the Norma arm tangent, and an interarm region. While the spectra do not show much variance between our models in all those specially motivated regions, for the flux projections we find local differences in the directions of the spiral arms.

All in all, our approach of combining observed and suitable simulated sources from a three-dimensional distribution makes our source model a novel enhancement for CR propagation.

For future modelling it would be beneficial to have a refined distinction between the different source classes, like SNRs and PWNe as the main candidates but also other astrophysical objects like $\gamma$-ray binaries by including their individual flux levels and source spectra.This would add a greater degree of accuracy to the model and the subsequent transport simulation. Furthermore,  improvements in new analytical methods of the observed data will be useful to better detect extended sources (see e.g. \citep{inproceedings}). Another important aspect is to further improve the underlying simulated source distribution (\citep{steppa_constantin_model_nodate}), e.g. by considering a central Galactic bar (\citep{2009PASP121213C}). This will influence our individual realisations since they depend on the used simulated source population. As mentioned above, the simulation of a time-dependent solution of CR transport, like in \citep{articlepjm}, will possibly also improve our results, since the sources are evolving with time. Finally, future instruments with increased sensitivity like CTA \cite{hermann_cherenkov_2007} will enhance the depth of Galactic coverage allowing to see a more complete picture of the Galactic $\gamma$-ray sources and resolve the diffuse $\gamma$-ray background more deeply.
This will lead to a enlarged observed source sample and, therefore, to refinements of source models which will allow us to study CR-propagation in the Galaxy more thoroughly. 
\section*{Acknowledgement}
We'd like to thank Katrin Egberts and Constantin Steppa for providing the simulated source sample as well as the many fruitful discussions. 
 \bibliographystyle{elsarticle-num} 
 \bibliography{test.bib}

\end{document}